\documentclass[twocolumn,showpacs,preprintnumbers,amsmath,amssymb]{revtex4}

\usepackage{graphicx}
\usepackage{dcolumn}
\usepackage{bm}

\newcommand{\bea}{\begin{eqnarray}}
\newcommand{\eea}{\end{eqnarray}}
\newcommand{\simgt}{\hbox{ \raise3pt\hbox to 0pt{$>$}\raise-3pt\hbox{$\sim$} }}
\newcommand{\simlt}{\hbox{ \raise3pt\hbox to 0pt{$<$}\raise-3pt\hbox{$\sim$} }}

\begin{document}

\preprint{TU--947}

\title{Perturbative heavy quarkonium spectrum at 
next-to-next-to-next-to-leading order}

\author{Y.~Kiyo$^1$ and Y.~Sumino$^2$\vspace*{3mm}
}
\affiliation{
$^1$Department of Physics, Juntendo University,
Inzai, Chiba 270-1695, Japan
\\
$^2$Department of Physics, Tohoku University,
Sendai, 980--8578 Japan
}%

\date{\today}

\begin{abstract}
We compute the energy levels of some of the lower-lying heavy quarkonium 
states perturbatively up to 
${\cal O}(\alpha_s^5 m)$ and ${\cal O}(\alpha_s^5 m \log\alpha_s )$.
Stability of the predictions
depends crucially on the unknown 4-loop pole-$\overline{\rm MS}$
mass relation.
We discuss the current status of the predictions
with respect to the observed bottomonium spectrum.
\end{abstract}

\maketitle

During the past decade, spectroscopy of heavy quarkonium states (in particular
the bottomonium states) has provided an important testing ground
of perturbative QCD.
On the one hand, we have 
at our disposal relatively many terms of the 
perturbative expansions.
On the other hand, the system in question is a small-size (compared to the typical 
QCD scale $\Lambda_{\rm QCD}^{-1}$) color-singlet object, from which
large part of infra-red (IR) degrees of freedom decouple.
In fact, the discovery of a cancellation of the  
IR renormalons in
the energy levels 
of heavy quarkonium states led to a drastic improvement in the
predictability of the energy levels
within perturbative QCD \cite{Pineda:id}.
We observe that
stability and convergence properties of the perturbative series
for the energy levels are fairly good, even in comparison to other
observables of a heavy quarkonium, such as production cross 
sections, transition rates, or partial decay widths.
Important applications of the spectroscopy are precise determinations
of the heavy quark masses from the lowest-lying energy levels.
The bottom and charm quark masses have been determined,
and in the future the top quark mass is expected to be determined
accurately in this way.
(See  \cite{Brambilla:2004wf} for reviews.)

The full ${\cal O}(\alpha_s^4 m)$ corrections to the energy levels 
were computed in \cite{Pineda:1997hz}.
Analyses of the bottomonium spectrum, which incorporate the
renormalon cancellation and the perturbative corrections up to
this order, have shown that the gross structure
of the bottomonium spectrum, including the levels of the
$n=1,2$ and some of the $n=3$ states
($n$ is the principal quantum number), is reproduced reasonably
well within the estimated perturbative uncertainties
\cite{Brambilla:2001fw}.

During the subsequent years,
our understanding
on the energy of a heavy quarkonium system based on
perturbative QCD has been advanced.
Stability and agreement with experimental data of the predictions for the
energy levels are predominantly determined by 
the prediction for the static QCD potential $V_{\rm QCD}(r)$.
After canceling the renormalon in $V_{\rm QCD}(r)$ (in various schemes),
perturbative predictability improves and the predictions for $V_{\rm QCD}(r)$
agree with lattice computations or typical phenomenological potentials
in the relevant distance range
\cite{Sumino:2001eh,Pineda:2002se}.
Furthermore, by increasing the order of the perturbative expansion, 
the range of $r$, where
convergence and agreement are seen, extends to
larger $r$ \cite{Anzai:2009tm}.
The details, however, depend on the schemes adopted for
canceling the renormalon.

Taking these features into account, some improvements of the predictions
for finer structures of the bottomonium spectrum have been examined.
Including all the known terms of $V_{\rm QCD}(r)$ in the zeroth-order
Hamiltonian, 
the fine and hyperfine splittings as well as the splittings between $S$-
and $P$-wave levels have been computed in a specific
organization of perturbative series
\cite{Recksiegel:2002za}.
This prescription enables to incorporate the effects of
the rise of $V_{\rm QCD}(r)$
at larger $r$
on the wave functions, and this
results in a better agreement of the above  
splittings with the experimental 
data.
(See also \cite{Kniehl:2003ap}.)
\footnote{
Although there were some
discrepancies between the experimental data and perturbative predictions
at earlier stages, newer experimental data are in good
agreement with the perturbative predictions
\cite{Skwarnicki:2003wn}.
}

In the meantime, computations of the ${\cal O}(\alpha_s^5 m )$
and ${\cal O}(\alpha_s^5 m \log\alpha_s )$ corrections
to the energy levels have made progress.
Development of 
effective field theories, such as
potential non-relativistic QCD (pNRQCD) \cite{Pineda:1997bj} or
velocity non-relativistic QCD \cite{Luke:1999kz}, 
enabled systematic computations of the higher-order corrections,
by separating the different kinematical regions involved
in the corrections.

Within pNRQCD, the corrections consist of two parts,
the contributions from the potential region and ultra-soft (US) region.
The next-to-next-to-next-to-leading order
(NNNLO) Hamiltonian, which dictates the
contributions from the potential region, was computed in \cite{Kniehl:2002br}
(besides the 3-loop corrections 
to $V_{\rm QCD}(r)$, $a_3$, which were computed later in
\cite{Smirnov:2008pn,Anzai:2009tm}).
It is a straightforward (but cumbersome) computation to obtain the
energy levels of the Hamiltonian in perturbative expansions
analytically.
The contributions from the US region contain, besides the
part calculable analytically, 
a QCD analogue of the Bethe logarithm for the Lamb shift in QED.
The QCD Bethe logarithm for each state can be written as a one-parameter integral of 
elementary functions \cite{Kniehl:1999ud}.
Up to now, the ${\cal O}(\alpha_s^5 m \log \alpha_s)$ correction for a general  state
labeled by the quantum numbers
$(n,l,s,j)$ was computed in \cite{Brambilla:1999xj}, while 
the ${\cal O}(\alpha_s^5 m)$ and ${\cal O}(\alpha_s^5 m \log \alpha_s)$
corrections for a general $S$-wave state  
$(n,j)$  were computed in \cite{Beneke:2005hg}.
We have confirmed these results.
(See also \cite{Kiyo:2000fr} for earlier computations
of the ${\cal O}(\alpha_s^5 m)$ 
corrections.)

\begin{table*}[t]
$$
\begin{array}{c|l}
\hline
(n,l) & \hspace*{60mm}c_3(n,l,s,j) \\
\hline
(1,0) & -0.447879 \, n_l^3+27.3508 \, n_l^2-418.003 \, n_l+597.111 \log\alpha_s+1928.76(1)+\mathbb{S}^2
   \left(-61.4109 \log\alpha_s-11.5278 \, n_l+218.589\right) \\
 (2,0) & -0.470041 \, n_l^3+29.0777 \, n_l^2-427.286 \, n_l+329.535 \log\alpha_s+1555.66(1)+\mathbb{S}^2
   \left(-30.7054 \log\alpha_s-10.7155 \, n_l+189.250\right) \\
 (2,1) & -0.413823 \, n_l^3+25.3451 \, n_l^2-414.351 \, n_l+108.748 \log\alpha_s+1968.47(1)
+\mathbb{S}^2 \left(0.162463
   \, n_l-0.121847\right)
\\&~
+D_S
   \left(-2.19325 \log\alpha_s-0.560973 \, n_l+13.1915\right)+ X_{LS}\left(-4.38649 \log\alpha_s-1.43923 \, n_l+41.1222\right)
   \\
 (3,0) & -0.454201 \, n_l^3+28.6079 \, n_l^2-418.477 \, n_l+236.444 \log\alpha_s+1419.35(1) +\mathbb{S}^2
   \left(-20.4703 \log\alpha_s-9.14505 \, n_l+158.960\right)\\
 (3,1) & -0.454469 \, n_l^3+27.7382 \, n_l^2-446.928 \, n_l+89.2529 \log\alpha_s+2035.04(1)
 +\mathbb{S}^2 \left(0.108308
   \, n_l-0.0812313\right)
\\&~
+D_S
   \left(-1.46216 \log\alpha_s-0.608358 \, n_l+12.5233\right)+ X_{LS}\left(-2.92433 \log\alpha_s-1.66261 \, n_l+38.7400\right)
   \\
 (3,2) & -0.400872 \, n_l^3+24.6125 \, n_l^2-402.879 \, n_l+69.7574 \log\alpha_s+1921.30(1)
+\mathbb{S}^2 \left(0.0216617
   \, n_l-0.0162463\right)
\\&~
+D_S
   \left(-0.292433 \log\alpha_s-0.0564700 \, n_l+1.81875\right)+ X_{LS}\left(-0.584865 \log\alpha_s-0.136917 \, n_l+5.30033\right)
   \\
 (4,1) & -0.468374 \, n_l^3+28.6896 \, n_l^2-459.027 \, n_l+78.7741 \log\alpha_s+2037.85(1)
+\mathbb{S}^2 \left(0.0812313
   \, n_l-0.0609235\right)
\\ &~
+D_S
   \left(-1.09662 \log\alpha_s-0.583921 \, n_l+11.4578\right)+
   X_{LS}\left(-2.19325 \log\alpha_s-1.62992 \, n_l+35.2919\right)
\\
\hline
\end{array}
$$
\vspace*{-3mm}
\caption{\label{results}
$c_3\equiv P_3(0)$ 
in the NNNLO
predictions for some of the  energy levels.
See text for the definitions of parameters.
}
\vspace*{-3mm}
\end{table*}

In this paper we present the results of our computation for the
${\cal O}(\alpha_s^5 m )$
and ${\cal O}(\alpha_s^5 m \log\alpha_s )$
corrections to the energy levels including some of the
$P$- and $D$-wave states.
Since the analytic expressions plus integral forms are too lengthy 
to be presented here, and since one-parameter
integrals need to be evaluated numerically for individual $(n,l)$'s in any case,
we present the results numerically for some $(n,l)$'s.
(The full formula and the derivation will be presented elsewhere.)
In particular, we present the corrections necessary for all the
observed bottomonium states whose masses are listed in \cite{Beringer:1900zz}
and which lie below the threshold for decays into two
$B$ mesons ($2M_B=10.558$~GeV).

We consider a bound-state composed of a 
quark (with the pole mass $m_{\rm pole}$) and its anti-quark.
The energy of the bound-state $X$, identified by $(n, l, s,j)$, is given by
\bea
&&
E_X(\mu,\alpha_s(\mu),m_{\rm pole})
\nonumber\\&&
=m_{{\rm pole}}\left[ 2
-\frac{C_F^{\, 2}\alpha_s(\mu)^2}{4n^2}\sum_{i=0}^\infty 
\biggl(\frac{\alpha_s(\mu)}{\pi}\biggr)^i \, P_i(L)
\right] ,
\label{MX}
\eea
with
\bea
&&
L = \log\left(\frac{n\mu}{C_F\alpha_s(\mu) m_{{\rm pole}}}\right)
+\sum_{k=1}^{n+l} \frac{1}{k} \, .
\eea
Here, 
$C_F=4/3$ denotes the color factor;
$\alpha_s(\mu)$ denotes the strong coupling constant
in the theory with $n_l$ active flavors only,
renormalized at the renormalization scale $\mu$, and
defined in the modified-minimal-subtraction ($\overline{\rm MS}$) scheme;
$P_i(L)$ denotes an $i$-th-degree polynomial of $L$.
$\alpha_s(\mu)$ 
obeys the renormalization-group (RG) equation
\bea
\mu^2 \, \frac{d}{d\mu^2} \, \alpha_s(\mu) =
- \alpha_s(\mu) \sum_{i=0}^{\infty} \beta_i
\biggl( \frac{\alpha_s(\mu)}{4\pi} \biggr)^{i+1},
\label{RGeq}
\eea
where $\beta_i$ represents the $(i+1)$-loop coefficient of the
beta function.
The only part of $P_i(L )$
that is not determined by the RG equation for $E_X$ is 
$c_i \equiv P_i(0)$.\footnote{
All the logarithms, which contain $\mu$ in the arguments, are rewritten in terms of $L$.
}
For $i= 3$, we have
\bea
&&
P_3=\frac{1}{2} \beta _0^3 L^3+
\left(-\frac{7 \beta _0^3}{8}+\frac{7 \beta _0 \beta _1}{16}+\frac{3}{2} \beta _0^2
   c_1\right) L^2
\nonumber\\&&
+\left(\frac{\beta _0^3}{4}-\frac{\beta_0 \beta _1 }{4}+\frac{\beta _2}{16}-\frac{3}{4} \beta _0^2 c_1+2 \beta _0 c_2+\frac{3
   \beta _1 c_1}{8}\right) L
\nonumber\\&&
 +c_3 .
 \label{c3}
\eea
Our results of $c_3$ are listed in Tab.~\ref{results}, given
as functions of $(s,j)$, $n_l$ and $\log[\alpha_s(\mu)]$ for fixed
$(n,l)$'s.\footnote{
The errors in numerics, shown by brackets, originate from
the error in $a_3$.
}
Here,
\bea
&&
\mathbb{S}^2\equiv \left< \vec{S}^2 \right> =s(s+1),
\\&&
X_{LS} \equiv
\left< \vec{L}\cdot \vec{S} \right>
= \frac{1}{2}\,
\left[ j(j+1)-l(l+1)-\mathbb{S}^2 \right] ,
\\&&
D_{S} \equiv
\left< 
3 {(\vec{r}\cdot \vec{S})^2}/{r^2} - \vec{S}^2 
\right>
\nonumber\\&&
~~~~
=
\frac{
2 l (l+1) \mathbb{S}^2 - 3 X_{LS} - 6 X_{LS}^2
}{
(2l-1)(2l+3)
} .
\eea
We neglect the masses of $n_l$ light quarks.
The non-logarithmic terms of the $P$- and $D$-wave levels are new.

\begin{figure*}[t]
\includegraphics[width=8cm]{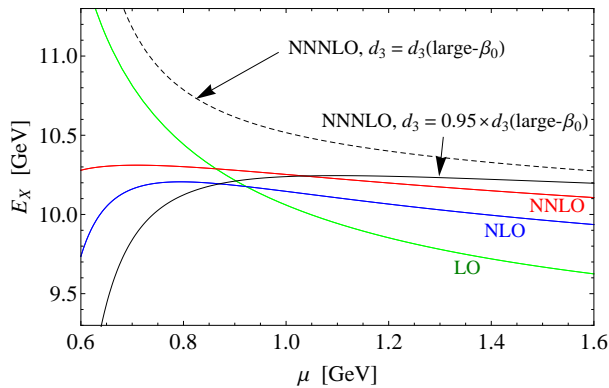}
\vspace*{-2mm}
\\
\caption{\small
$E_X$ for $\chi_b(2^3\!P_0)$ 
as a function of $\mu$.
The solid lines represent the sum of the
perturbative series up to 
${\cal O}(\alpha_s^2m)$ [LO], ${\cal O}(\alpha_s^3m)$ [NLO],
${\cal O}(\alpha_s^4m)$ [NNLO] and 
${\cal O}(\alpha_s^5 m,\alpha_s^5 m\log\alpha_s)$ [NNNLO,
$d_3=0.95\,d_3^{\mbox{\scriptsize large-}\beta_0}$].
The dashed line represents the
NNNLO prediction with
$d_3=d_3^{\mbox{\scriptsize large-}\beta_0}$.
The $\varepsilon$-expansion is used for canceling the renormalons.
\label{FigScaledep}
}
\end{figure*}

\begin{table*}[t]
\vspace*{-3mm}
{\small
\bea
\begin{array}{ccrrrrrrll}
\hline
X  & (n,l,s,j) & E_X^{\rm exp} & E_X^{\rm pert}  & E_X^{(1)} & E_X^{(2)} & E_X^{(3)} & E_X^{(4)} & \mu_X & 
\mbox{\hbox{\hbox to 20pt{$\!\!\alpha_{\!s}(\mu_{X}\!)$}}}
\\ \hline
 \eta _b{(1^1\! S_0)}  & (1,0,0,0) & 9.398 & 9.441 & 0.64 & 0.26 & 0.10 & 0.013 & 6.26 & 0.199 \vspace*{1mm}\\
 {\Upsilon(1^3\! S_1)}  & (1,0,1,1) & 9.460 & 9.460 & 0.67 & 0.26 & 0.10 & 0.011 & 5.38 & 0.209 \vspace*{1mm}\\
 \chi _{{b}}{(1^3\! P_0)}  & (2,1,1,0) & 9.859 & 9.893 & 1.06 & 0.29 & 0.11 & 0.007 & 1.95 & 0.308 \vspace*{1mm}\\
 \chi _{{b}}{(1^3\! P_1)}  & (2,1,1,1) & 9.893 & 9.900  & 1.07 & 0.29 & 0.11 & 0.007 & 1.90 & 0.313 \vspace*{1mm}\\
 h_b{(1^1\! P_1)}  & (2,1,0,1) & 9.899 & 9.902  & 1.08 & 0.28 & 0.11 & 0.007 & 1.88 & 0.314 \vspace*{1mm}\\
 \chi _{{b}}{(1^3\! P_2)}  & (2,1,1,2) & 9.912 & 9.905  & 1.09 & 0.28 & 0.11 & 0.006 & 1.85 & 0.317 \vspace*{1mm}\\
 \eta _b{(2^1\! S_0)}  & (2,0,0,0) & 9.999 & 9.951  & 1.13 & 0.28 & 0.10 & 0.009 & 1.69 & 0.332 \vspace*{1mm}\\
 {\Upsilon(2^3\! S_1)}  & (2,0,1,1) & 10.023 & 9.962 & 1.15 & 0.27 & 0.11 & 0.010 & 1.66 & 0.335 \vspace*{1mm}\\
 {\Upsilon(1^3\! D_2)}  & (3,2,1,2) & 10.164 & 10.180  & 1.41 & 0.22 & 0.12 & 0.014 & 1.22 & 0.403 \vspace*{1mm}\\
 \chi _{{b}}{(2^3\! P_0)}  & (3,1,1,0) & 10.233 & 10.245 & 1.52 & 0.16 & 0.12 & 0.019 & 1.10 & 0.435 \vspace*{1mm}\\
 \chi _{{b}}{(2^3\! P_1)}  & (3,1,1,1) & 10.255 & 10.253 & 1.54 & 0.15 & 0.12 & 0.020 & 1.08 & 0.441 \vspace*{1mm}\\
 h_b{(2^1\! P_1)}  & (3,1,0,1) & 10.260 & 10.256  & 1.54 & 0.15 & 0.12 & 0.020 & 1.07 & 0.443 \vspace*{1mm}\\
 \chi _{{b}}{(2^3\! P_2)}  & (3,1,1,2) & 10.269 & 10.259 & 1.55 & 0.14 & 0.12 & 0.021 & 1.07 & 0.445 \vspace*{1mm}\\
 {\Upsilon(3^3\! S_1)}  & (3,0,1,1) & 10.355 & 10.324 & 1.65 & 0.09 & 0.13 & 0.029 & 0.98 & 0.475 \vspace*{1mm}\\
 \overline{\chi_{\!b}{(3^3\! P_{\!j})}} & (4,1,1,j_{\rm av}) & 10.534 & 10.692 & 2.21 & 
\mbox{\hbox{\hbox to 4pt{$\!\!-$}}}0.31 
& 0.30 & 0.068 & 0.75 & 0.632
\\
\hline
\end{array}
\nonumber
\eea
}
\vspace*{-4mm}
\caption{\label{tab2}
Experimental values vs.\ perturbative predictions for $E_X$
in the case $d_3=0.95\,d_3^{\mbox{\scriptsize large-}\beta_0}$.
$E_X^{(i)}$ denotes the $i$-th order term of the $\varepsilon$-expansion.
$E_X^{\rm pert}=2\overline{m}+\sum_{i=1}^4 E_X^{(i)}$.
Numerical values except in the second and last columns are in GeV.
The last row represents 
the spin-averaged $3P_j$ energy for $j=0,1,2$
with the weight factor $2j +1$.
}
\vspace*{-1mm}
\end{table*}

Using the NNNLO results we compute the energies of the
observed bottomonium states and compare them with the experimental data.
We follow the prescription used in the analyses \cite{Brambilla:2001fw}.\footnote{
Bottomonium $S$-state levels at NNNLO 
have been examined in different schemes \cite{Beneke:2005hg}.
At NNLO it is known that the scheme of \cite{Brambilla:2001fw} gives
an optimal convergence behavior.
}
To cancel the renormalons, we express the pole mass in terms of
the quark mass defined in the $\overline{\rm MS}$ scheme ($\overline{\rm MS}$ mass)
as
\bea
m_{\rm pole}=\overline{m}\left[ 1 + \sum_{i=0}^{\infty}
\left(\frac{\alpha_s(\overline{m})}{\pi}\right)^{i+1} d_{i}
\right] ,
\label{pole-mass}
\eea
where $\overline{m} \equiv m_{\overline{\rm MS}}(m_{\overline{\rm MS}})$
denotes the $\overline{\rm MS}$ mass renormalized at the
$\overline{\rm MS}$ mass scale.
The 4-loop constant $d_3$, which is needed for our analysis, is not known yet.
Up to now there exist some estimates of $d_3$ \cite{bb,pinedaJHEP,Kataev:2010zh,SuminoEst}.
We adopt the estimate \cite{SuminoEst},
obtained from perturbative stability of the energy of a static quark pair
in the following manner.
The upper bound of the estimate is 
determined by requiring stability of the perturbative prediction
for $E_{\rm tot}(r)\equiv 2m_{{\rm pole}}+V_{\rm QCD}(r)$  at NNNLO
at least up to the same $r$ as NNLO.
In particular, as the value of $d_3$ exceeds its estimated upper bound, the 
perturbative prediction for
$E_{\rm tot}(r)$ becomes unstable rapidly.
The lower bound of the estimate is obtained by requiring that the difference
between the NNLO and NNNLO predictions for $E_{\rm tot}(r)$ be within
an  ${\cal O}(\Lambda_{\rm QCD}^3r^2)$ uncertainty.
When $d_3$ is chosen within the estimated range, qualitatively the
prediction for $E_{\rm tot}(r)$ becomes stable and the series exhibits a
reasonably convergent behavior.

After rewriting $E_X$ in terms of $\overline{m}$ and
$\alpha_s(\mu)$ via eq.~(\ref{pole-mass})
and the solution to eq.~(\ref{RGeq}), we expand $E_X$ 
in $\alpha_s(\mu)$. 
To make the cancellation of the renormalons explicit, we 
reorder the series in the
so-called ``$\varepsilon$-expansion scheme'' \cite{Hoang:1998ng}.
See \cite{Brambilla:2001fw} for details.
We set $\alpha_s(M_Z)=0.1184$, $n_l=4$, and $\overline{m}$
is fixed such that the energy of the $\Upsilon (1^3\!S_1)$ state coincides
with the experimental value.
We vary the value of $d_3$ within
$(0.95^{+ 0.01}_{-0.05})\! \times \! d_3^{\mbox{\scriptsize large-}\beta_0}$,
which is the stability range of 
$E_{\rm tot}(r)$\footnote{
This range of $d_3$ corresponds to 
$\overline{m}\sim 4$~GeV.
The range of $d_3$ for large $\overline{m}$ is 
$(0.95^{+ 0.02}_{-0.05})\! \times \! d_3^{\mbox{\scriptsize large-}\beta_0}$
\cite{SuminoEst}.
},
where $d_3^{\mbox{\scriptsize large-}\beta_0}\approx 1324.49$
for $n_l=4$ \cite{bb}.\footnote{
For comparison,
the estimates by renormalon dominance
\cite{pinedaJHEP} give
$d_3\approx (0.99$--$1.02)\!\times\! d_3^{\mbox{\scriptsize large-}\beta_0}$,
while the estimate \cite{Kataev:2010zh} gives 
$d_3\approx 0.74\, d_3^{\mbox{\scriptsize large-}\beta_0}$,
for $n_l=4$.
}
We find that practically the stability of the perturbative prediction for
$E_X$ is determined by the stability of the perturbative prediction
for $E_{\rm tot}(r)$.
In fact, the scale dependence and convergence properties
of $E_X$ are qualitatively similar to those of $E_{\rm tot}(r)$.
In Fig.~\ref{FigScaledep} we show the scale dependence of $E_X$ for the
$\chi_b(2^3\!P_0)$ state, which is one of the highest states 
predicted reliably at NNLO.
The NNNLO prediction is stable if $d_3$ is within the above range.
If $d_3$ is raised above
$0.96\, d_3^{\mbox{\scriptsize large-}\beta_0}$, 
the NNNLO prediction becomes unstable quickly,
while
if $d_3$ is reduced below 
$0.90\, d_3^{\mbox{\scriptsize large-}\beta_0}$, 
convergence and stability of the prediction
become worse gradually.

\begin{figure*}[t]
\includegraphics[width=8cm]{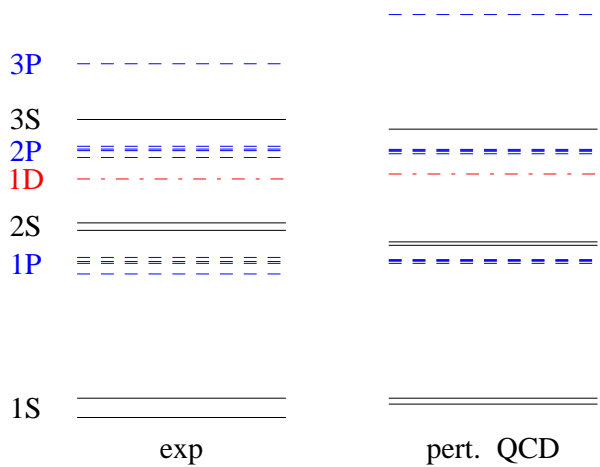}
\vspace*{-1mm}
\\
\caption{\small
Bottomonium spectrum as given by
experiments and by our analysis in Tab.~\ref{tab2}.
The solid, dashed, and dotdashed lines represent, respectively,
$S$-, $P$- and $D$-wave levels.
There are four lines for the $1P_j$ and $2P_j$ states
(spin triplet and singlet states), respectively, while
only one line is shown for the $1D$ state
corresponding to 
the $(s,j)=(1,2)$ state.
\label{FigLineSpect}
}
\vspace*{-1mm}
\end{figure*}

Let us fix $d_3$ to 
$0.95\, d_3^{\mbox{\scriptsize large-}\beta_0}$, an optimal value
with respect to the
stability of $E_{\rm tot}(r)$.
Then, for each state $X$, 
we fix the scale $\mu=\mu_X$ by demanding stability of 
$E_X$ against variation of the scale:
\bea
\mu \frac{d}{d\mu}E_X(\mu,\alpha_s(\mu),\overline{m})\biggr|_{\mu=\mu_X}=0 .
\label{scaleX}
\eea
This scale exists for all the bottomonium states considered here.
The convergence behaviors of the perturbative 
expansions are reasonable.
This means that the predictability range extends to
higher levels compared to the NNLO predictions.
We list the perturbative predictions and the experimental
data in Tab.~\ref{tab2}.\footnote{
We neglect errors of the experimental data, which are much
smaller than errors of the perturbative predictions.
}
The bottom quark $\overline{\rm MS}$ mass, fixed on the
$\Upsilon (1^3\!S_1)$ state, is given by\footnote{
If the value of $\overline{m}$ is varied by $\Delta \overline{m}$, 
all the energy levels $E_X$
are shifted approximately by $2\Delta \overline{m}$
such that all the level spacings $E_X-E_{X'}$
are barely changed.
}
\bea
\overline{m}\equiv m_b^{\overline{\rm MS}}(m_b^{\overline{\rm MS}})
=4.214~\mbox{GeV} .
\eea
In Fig.~\ref{FigLineSpect} we compare the experimental data and the
predicted bottomonium spectrum.
We see a reasonable agreement for the gross structure of
the spectrum.
(See, however, the comments on uncertainties below.)

If $d_3$ is within the range $(0.90$--$0.96)\! \times \! d_3^{\mbox{\scriptsize large-}\beta_0}$, 
convergence properties and stability of the 
predictions are qualitatively similar to those listed
in Tab.~\ref{tab2}, although the level of agreement with the experimental
data varies.
As we raise (reduce) the value of $d_3$, 
level spacings among different states increase (decrease).
Variations are larger for higher states.
If $d_3$ is raised above $0.96\, d_3^{\mbox{\scriptsize large-}\beta_0}$,
the extremum scale $\mu_X$ disappears quickly from higher levels.
If 
$d_3>1.2\, d_3^{\mbox{\scriptsize large-}\beta_0}$,
$\mu_X$ no longer exists
even for the $\Upsilon(1^3\!S_1)$ state.

In principle, we can estimate uncertainties of the predictions for $E_X$ 
originating
from various sources, 
similarly to the analyses \cite{Brambilla:2001fw},  for each 
given value of $d_3$.
With respect to the uncertainties, 
we can  discuss agreement or disagreement with the experimental data.
It would eventually lead to,
for instance, quantification of non-perturbative contributions
to individual energy levels.
Since, however, stability of the predictions for $E_X$  depends 
crucially on $d_3$,
we consider the current status to be too premature to do this quantitatively.
Here, we briefly comment on uncertainties.
\\
(i) {\it Dependence on $\alpha_s(M_Z)$}:
If we vary $\alpha_s(M_Z)$ within the current uncertainties
$\pm 0.0007$ \cite{Beringer:1900zz}, variation of the energy levels
[after fixing $\overline{m}$ on $\Upsilon(1^3\! S_1)$]
is fairly small and minor as compared to uncertainties from
other sources.
\\
(ii) {\it Non-zero charm mass effects}:
Although a full account of non-zero charm mass effects in loops
requires a separate analysis of its own, the analysis at NNLO indicates
that the level spacings among higher levels increase due to
the decoupling of the charm quark.
Phenomenologically this indicates that a smaller value of $d_3$
may be favored for a better agreement with the experimental data
after inclusion of these effects.
\\
(iii) {\it Higher-order effects on level splittings}:
As already explained, inclusion of higher-order effects 
of $V_{\rm QCD}(r)$ in
the bound-state wave functions increases the fine and hyperfine splittings
as well as the $S$-$P$ splittings, which improves
the agreement with the experimental
data.
We note that concerning the former splittings all the NNNLO corrections included 
in the present analysis have already been included in \cite{Recksiegel:2002za},
so that the differences from the our results 
stem only from higher-order effects.
\\
(iv) {\it Scale dependences}:
If $d_3$ is within the range 
$(0.90$--$0.96)\! \times \! d_3^{\mbox{\scriptsize large-}\beta_0}$, 
the scale dependences of $E_X$ are reduced as compared
to the NNLO predictions.
For instance, in the case
$d_3=0.95\, d_3^{\mbox{\scriptsize large-}\beta_0}$
if we choose the scale $\mu=2\mu_X$, $E_X$ varies by
30--50~MeV for the $n=2$ levels, by 80--120~MeV
for the $n=3$ levels, and by 250--300~MeV for the $n=4$ levels.
(The scales $\mu=\mu_X/2$ are too low to give sensible
predictions.)

Let us comment on
non-perturbative contributions to the bottomonium energy levels.
In general there are two ways to compute a physical quantity whose
major contributions come from UV region.
One way is to compute thoroughly within perturbative QCD.
The other way is to compute by factorizing UV and IR contributions
within a Wilsonian low energy effective theory.
In the former computation, there are well-established prescriptions to estimate
uncertainties of the prediction within perturbative QCD.
Empirically estimates of 
perturbative uncertainties are approximated well by IR renormalons,
in the case that IR renormalons turn out to be large.
In the latter computation, UV contributions are encoded in the Wilson 
coefficients, which are free from IR renormalons and have small 
uncertainties once higher-order corrections are known, while IR contributions
are included in non-perturbative matrix elements.
The correspondence of the two computations is that IR part of the former 
computation is replaced by the matrix elements of the latter, and that the 
residual UV contributions of the former equals the Wilson coefficients of
the latter.
Thus, the uncertainties by IR renormalons in the former computation are
replaced by the non-perturbative matrix elements in the latter computation.
Our computation in this paper corresponds to the former type of computation. 
A meaningful and 
consistent comparison between the two types of computations 
would be to compare perturbative uncertainties (IR 
renormalons) with direct evaluation of the leading non-perturbative matrix 
elements.\footnote{
To our knowledge there is no theoretical formulation which justifies
to simply add non-perturbative contributions of the latter
type of computation to the former type of computation.
}

We have computed the quarkonium
energy levels perturbatively.
In particular, the US corrections, which first appear
at NNNLO, are computed 
perturbatively.
IR contributions from the scale
of order $\Lambda_{\rm QCD}$ in
these computations give rise to uncertainties 
(IR renormalons) of order $\Lambda_{\rm QCD}^3 a_X^2$.
($a_X$ denotes the typical size of the quarkonium state $X$.)
The above estimates (iv) of perturbative uncertainties
of our predictions are consistent with this estimate.
Within pNRQCD, 
non-perturbative (IR) contributions and
UV contributions
can be factorized \cite{Pineda:1997bj}.
The former are given by
matrix-elements of non-local gluon condensates;
the latter are given by the
Wilson coefficients, which can be predicted reliably by
perturbative QCD, free from IR renormalons. 
The leading-order non-perturbative contribution 
is estimated to be of order $\Lambda_{\rm QCD}^3 a_X^2$ from
dimensional analysis.
Thus, the renormalon uncertainty can be absorbed into
a non-perturbative matrix element with the same order of magnitude.\footnote{
This type of relations between IR renormalons
(perturbative uncertainties) and non-perturbative effects
appear in various observables
of QCD.
}
The analyses in \cite{Brambilla:2001fw,Recksiegel:2002za} confirm 
consistency of the
bottomonium spectrum at NNLO with this relation. 
A similar feature is confirmed also
for the static potential at NNLO and at NNNLO
in \cite{Pineda:2002se}.
Namely, 
it has been confirmed that the magnitudes of
perturbative uncertainties are of 
order $\Lambda_{\rm QCD}^3 a_X^2$ or
$\Lambda_{\rm QCD}^3 r^2$,
and that the perturbative predictions 
are consistent with the experimental data or with lattice
computations within the estimated
uncertainties.
Unfortunately, up to now, there exists no direct evaluation of
the leading-order non-local gluon condensate by lattice simulations
or by other methods.
A qualitatively new aspect of our present analysis consists in
the perturbative evaluation of the US corrections, which
includes the perturbative evaluation of
the non-local gluon condensates.
The convergence of the perturbative expansions
of the energy levels (within order $\Lambda_{\rm QCD}^3 a_X^2$
uncertainties)
observed at NNNLO seems to indicate that 
the perturbative evaluation of the gluon condensates provides
reasonable order-of-magnitude estimates
$\sim \Lambda_{\rm QCD}^3 a_X^2$.
However, a definite conclusion cannot be drawn until we know the
precise value of $d_3$.

Perhaps a well-known estimate of non-perturbative
contributions to the energy levels of heavy quarkonium states
is the Voloshin-Leutweyler formula expressed in terms of
the local gluon condensate 
$\sim n^6\langle \alpha_s G_{\mu\nu}^a(0)G_{\mu\nu}^a(0)\rangle
/(m^3\alpha_s^4)$
\cite{VL}.
As shown in \cite{Pineda:1997bj}, the non-local gluon condensates in pNRQCD
can be expressed by the local gluon 
condensates in the case that the 
time scale  of US gluons $T_{\rm US}\sim a_X/(C_A \alpha_s)$ is 
much smaller
than $1/\Lambda_{\rm QCD}$,
namely, in the case that $a_X$ is extremely small 
($\ll C_A \alpha_s/\Lambda_{\rm QCD}$).
If, in addition, the wave functions of the quarkonium states can be approximated 
by the Coulomb wave functions, we obtain the non-perturbative contributions as 
given by the Voloshin-Leutweyler formula.
Neither of these conditions, however, are met by the bottomonium states, especially 
by the excited states.
As shown by series of studies on heavy quarkonium states in perturbative QCD,
the bottomonium states lie in the intermediate-distance region, where deviation
of the static potential from the Coulomb potential by the higher-order QCD 
corrections is essential and where the US time
scale is not very much smaller than 
$1/\Lambda_{\rm QCD}$.
The Voloshin-Leutweyler formula is theoretically interesting but applicable only
to hypothetical ultra-heavy quarkonium states, which lie in a deep part of the 
Coulomb potential.
Inapplicability of the formula to the bottomonium states is signaled by an 
uncontrollably rapid increase (proportional to $n^6$) of the formula with the 
principal quantum number $n$.  
In fact, already for $n\sim 2$ -- 4, the formula gives numerically 
unrealistically large contributions.
This $n^6$ behavior results from a combination of (1) $r^2 T_{\rm US}\sim a_X^3$ 
behavior of the coefficient of the
local gluon condensate
(in contrast to $r^2\sim a_X^2$ 
behavior of the non-local condensate) and 
(2) a rapid increase
of the radius of the Coulomb state with $n$,  $a_X \propto n^2$,
since the potential becomes flat as $r$ 
increases;
note that,
it is the remediation of this
unphysical behavior of the potential that has been essential in reproducing 
the gross structure of the bottomonium energy levels within perturbative QCD.
Thus, such a rapid $n$-dependence cannot appear for the bottomonium states.

The scales $\mu_X$ for the $n\geq 2$ states
listed in Tab.~\ref{tab2} are small
and the corresponding
values of $\alpha_s(\mu_X)$ are large.
Hence, one may question validity of the
perturbative predictions.
Generally 
validity of a perturbative QCD prediction is examined through
stability against variation of scales,
convergence of perturbative series,
comparison with lattice computations,
and ultimately comparison with
the experimental data.
A common feature observed today in various (well-behaved) observables
of perturbative QCD is as follows.
The range of the energy scale where a prediction is stable
becomes wider as we include higher-order terms
of the perturbative series.
The range extends not only in the UV direction but also in
the IR direction.
The level of stability and convergence of perturbative predictions
depend on the observables and the typical scales involved in the observables.
The stability range of the static potential and (consequently)
that of the spectrum,
after cancellation of the leading-order renormalons,
have extended to surprisingly long-distance (IR)
region and higher states, respectively.
Concerning limitation of these perturbative predictions,
we believe that the predictions are evidently 
invalid at $r \simgt \Lambda_{\overline{\rm MS}}^{-1}\approx 1$~fm,
where the string-breaking phenomenon takes place, and
equivalently, above the $B\bar{B}$
threshold in the case of the energy levels.
On the other hand, in order to judge at which $r$ or the
energy level the
perturbative predictions break down before entering this non-perturbative
regime,
we have no other criteria than to
apply the above general prescriptions to examine validity
of the perturbative predicitons.
In this analysis we have presented a first examination of the
entire bottomonium spectrum at NNNLO.

The current status of the perturbative prediction for the
bottomonium spectrum
may be summarized as follows.
We expect that stability of 
$E_{\rm tot}(r)=2m_{\rm pole}+V_{\rm QCD}(r)$ is realized,
to a certain extent, as a result of decoupling of IR contributions
due to a general property of the gauge theory.
Nevertheless, the present status of the perturbative prediction
for the bottomonium spectrum
is practically determined by a fine-level cancellation
between $2m_{\rm pole}$ and $V_{\rm QCD}(r)$
and depends sensitively on the precise value of $d_3$.
If $d_3$ is tuned to stabilize $E_{\rm tot}(r)$
optimally,
we observe a reasonable agreement between the predictions 
and experimental data within estimated perturbative
uncertainties.

\section*{Acknowledgments}
The work of Y.S.\ is supported in part by 
Grant-in-Aid for
scientific research (No.\ 23540281) from
MEXT, Japan.


\end{document}